\documentclass[prb,twocolumn,superscriptaddress,eqsecnum]{revtex4}
\usepackage{bm}
\usepackage{graphicx}
\usepackage{amsmath}
\usepackage{amstext}
\usepackage{amssymb}
\usepackage{amsfonts}
\usepackage{amsbsy}

\begin{document}
\title{Cuprates as doped $U(1)$ spin liquids}
\author{T. Senthil}
\affiliation{Department of Physics, Massachusetts Institute of
Technology, Cambridge, Massachusetts 02139}
\author{Patrick A. Lee}
\affiliation{Department of Physics, Massachusetts Institute of
Technology, Cambridge, Massachusetts 02139}

\begin{abstract}

We explore theoretically the notion that the underdoped cuprates may be viewed as doped $U(1)$ spin liquid Mott
insulators. We pursue a conceptually clear version of this idea that naturally incorporates 
several aspects of the phenomenology of the cuprates. We argue that the low doping region may be fruitfully discussed in terms of 
the universal physics associated with a chemical potential tuned Mott transition between a $U(1)$ spin liquid insulator and a $d$-wave superconductor. 
A precise characterization of the deconfinement in the 
$U(1)$ spin liquid is provided by the emergence of a conserved gauge flux. This extra conservation law 
should hold at least approximately in the underdoped materials. Experiments that could possibly detect this conserved gauge flux are proposed.

\end{abstract}

\maketitle

\section{Introduction}

Superconductivity in the cuprate materials occurs upon doping Mott insulators. The evolution
of the physical properties
from the insulator to the superconductor as a function of doping has been extensively studied in the last several
years. Many experiments have clearly established the relevance of the proximity to the Mott state at
zero doping for understanding
phenomena at finite doping where superconductivity appears at low temperature\cite{ormil}.

\begin{figure}[tb]
\centering
\includegraphics[width=3.0in]{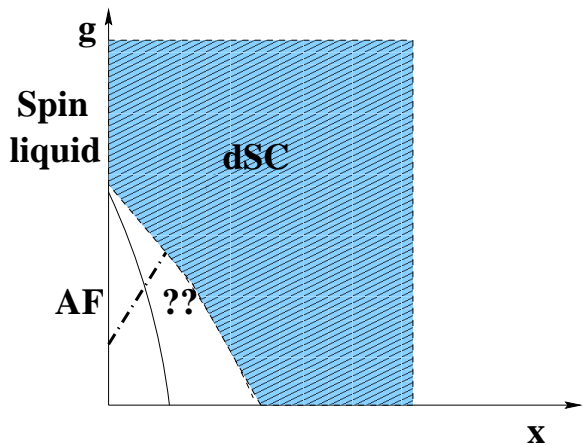}
\caption{Schematic zero temperature phase diagram showing the route between the antiferromagnetic Mott insulator and the $d$-wave superconductor. The vertical axis
is labelled by a parameter $g$ which may be taken as a measure of the frustration in the interaction between the spins in the Mott insulator. 
AF represents the antiferomagnetically ordered state. SL is a spin liquid insulator that could potentially be reached by increasing the frustration. 
The path taken by the cuprate materials as a function of doping $x$ is shown in a thick  dashed-dot line. The question marks represent regions where the physics is 
not clear at present. Doping the spin liquid naturally leads to the dSC state. The idea behind the spin liquid approach is to regard the superconducting system
at non-zero $x$  
as resulting from doping the spin liquid though this is not the path actually taken by the material.} \label{sldsc1}
\end{figure}

There is considerable theoretical debate
however on the precise nature of the connection between the undoped Mott state and the doped superconductor.
Experimentally it is well established that the Mott insulating state at zero doping develops
long ranged Neel antiferromagnetism.
Upon doping the magnetism disappears
rather rapidly and soon thereafter is replaced by the superconducting ground state. The ``normal'' state above the superconducting transition temperature
is metallic and is commonly referred to as the pseudogap state. A key observation is that the pseudogap state and the superconductor that descends from it
both remember the Mottness of their parent undoped material but do not remember its long ranged Neel order. 
This observation is encoded in the old suggestion\cite{pwa,subirrmp} that a
useful way to think about the underdoped region is to first
understand the nature of the possible non-magnetic Mott states at zero doping - obtained for instance by increasing frustration
in the interaction between the spins. The behaviour
of the doped system may then possibly be fruitfully viewed as the result of doping this non-magnetic Mott state.
The idea is that the doping effectively
frustrates the Neel order so that the system is pushed across the transition where the Neel order is lost (see Fig. \ref{sldsc1}).

In this paper we pursue a particular non-magnetic Mott state that is connected to the
Neel state by a second order transition. Furthermore doping this
non-magnetic state leads to a $d$-wave superconductor with nodal quasiparticle excitations.
For these and other reasons discussed below, the particular possibility
we explore is theoretically very appealing. We show that this
route to superconductivity leads to very unusual (and hence possibly unique)
physical effects that could possibly be detected in experiments.

The particular non-magnetic state we consider may be dubbed a `deconfined' $U(1)$ spin
liquid on the two dimensional square lattice. The low energy theory of this state
consists of nodal linear dispersing Dirac spinons that are coupled to a {\em non-compact}
fluctuating $U(1)$ gauge field. Such a spin liquid state has played a central role
in a large number of prior theoretical papers in the field\cite{affleck-marston-rapid,wlsu2,KL9930,RWspin}. Recent developments\cite{stable-u1} 
have considerably clarified the conceptual basis of this state opening the way to the
perspective used in this paper. Indeed the present paper may be viewed as providing
a conceptually clear version of various previous theoretical ideas that allows
for predictions that are possibly testable. 

\begin{figure}[tb]
\centering
\includegraphics[width=3.0in]{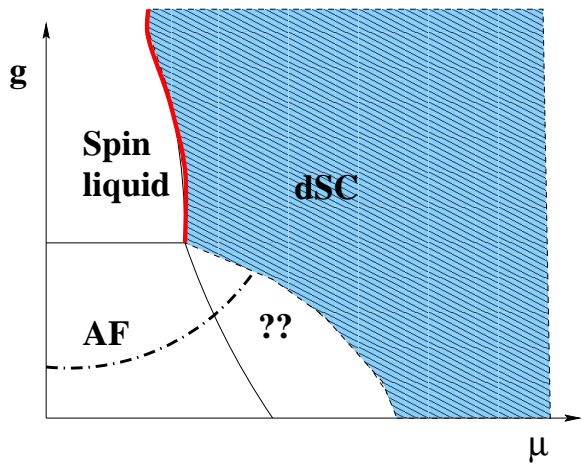}
\caption{Same as in Fig. \protect\ref{sldsc1} but as a function of chemical potential rather than hole doping.} \label{sldsc2}
\end{figure}

\begin{figure}[tb]
\centering
\includegraphics[width=3.0in]{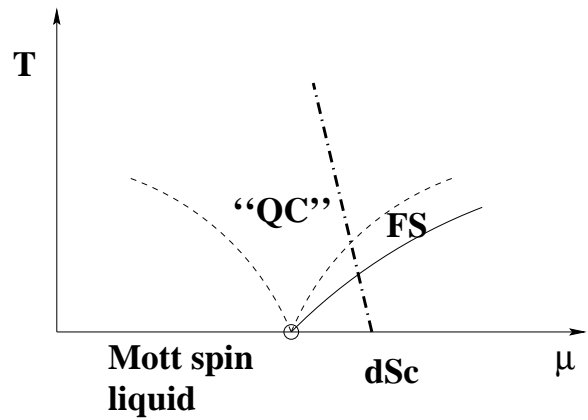}
\caption{Schematic phase diagram for a doping induced Mott transition between a spin liquid insulator and a $d$-wave superconductor. 
The bold dot-dashed line is the path taken by a system at hole density $x$ that has a superconducting ground state. The region marked FS represents the fluctuation regime
of the superconducting transition. The region marked QC is the quantum critical region associated with the Mott critical point. This region may be identified
with the high temperature pseudogap phase in the experiments. 
 } \label{mqc}
\end{figure}

What precisely is meant by the notion that the underdoped cuprates may be viewed as doped versions of a non-magnetic Mott insulator? 
To understand this it is instructive to consider the phase diagram as a function of the chemical potential rather than the hole doping as shown in Fig. \ref{sldsc2}. 
Consider any non-magnetic Mott state that when doped leads to a $d$-wave superconductor. As a function of chemical potential, there will then be a 
zero temperature phase transition where the holes first enter the system. For concretenes we will simply refer to this 
as the Mott transition - the corresponding phase boundary is marked in red in Fig \ref{sldsc2}. The associated quantum critical fixed point will control 
the physics in a finite non-zero range of parameters. The various crossovers expected near such transitions are well-known and are shown in Fig. \ref{mqc} 
Sufficiently close to this zero temperature critical point many aspects of the physics will be universal. The regime in which such universal behavior 
is observed will be limited by `cut-offs' determined by microscopic parameters. In particular we may expect that the cutoff scale is provided by an energy of 
a fraction of $J$ (the exchange energy for the spins in the Mott insulator). We note that this corresponds to a reasonably high temperature scale. 

Now consider an underdoped cuprate material at fixed doping $x$. Upon increasing the temperature this will follow a path in Fig.\ref{mqc} that is shown schematically. 
The properties of the system along this path may be usefully discussed in terms of the various crossover regimes in Fig. 
In particular it is clear that the `normal' state above the superconducting transition is to be understood directly as the finite temperature `quantum critical'
region associated with the Mott transition. Empirically this region corresponds to the pseudogap regime. Thus our assertion is that the pseudogap regime 
is controlled by the unstable zero temperature fixed point associated with the (Mott) transition to a Mott insulator.

Once we adopt this point of view it is clear that the Mott transition (which controls the pseudogap regime) need not be 
to the antiferromagnetic Mott state realized in the parent materials. Indeed the physical system takes a path through a complicated 
region which may involve charge ordering or stripe phases whose relation to other regions of the phase diagram is poor;y understood. 
Rather for the reasons mentioned in previous paragraphs it makes sense to explore 
Mott transitions to insulators that are non-magnetic. 
In the bulk of this paper we will focus on the $U(1)$ spin liquid insulator and the Mott transition from it to a superconductor induced by doping. 
Later we show how many aspects of the resulting theory may still survive provided only certain weaker assumptions are satisfied (see Section \ref{alternate}).

\section{Mott insulator}
We begin with a discussion of the Mott insulator. The Neel state is the ground
state of the nearest neighbour Heisenberg antiferromagnet. However
as explained above we will be interested
in possible non-magnetic ground states that are proximate to the Neel state.
It is very important to realize that there are very few such known candidate non-magnetic ground states. 
Thus this theoretical approach to the cuprate problem is fairly tightly constrained. 
 One
natural candidate is the dimerized state described in Ref. \onlinecite{ReSaSuN}.
Studies of the doped dimerized state have been pursued with some phenemenological success\cite{SaVoj}.
Indeed a superconducting state with $d$-wave symmetry
obtains upon doping. However such a superconductor also inherits the dimer order of the parent Mott state. 
In particular it has a full gap to spin excitations (at least at low doping) and breaks translation symmetry.
Empirically however there is strong evidence for the
presence of such nodal quasiparticles, and for the presence of translation symmetry in most of the materials studied. 
We are thus naturally led to search for 
translation invariant non-magnetic
Mott states that when doped will produce a $d$-wave superconductor
with nodal quasiparticles.

Mott insulators that preserve translation and other lattice symmetries are rather exotic beasts - the excitation spectra of all known 
theoretical examples are conveniently described in terms of fractionalized spin-$1/2$ `spinon' degrees of freedom. Further in certain 
such `spin liquid' states, it is possible for the spinons to have gapless nodal points with linear dispersion. Such spin liquids 
therefore provide ideal starting points to dope to produce translation invariant $d$-wave superconductors with nodal quasiparticles. 

An important physical property of such fractionalized spin liquid Mott insulators is the emergence of extra topological structure. 
Specifically spin liquid states possess extra topological conservation laws not present in the microscopic models in which they arise. 
This conservation law is conveniently interpreted as the flux of an emergent gauge field. Different classes of spin liquids may be defined 
depending on the nature of this conserved gauge flux.

A simple example of a spin liquid with nodal fermionic spinons is provided by a state where there is a conserved $Z_2$ gauge flux. Indeed an
attractive
theory of the cuprates based on doping this state was developed and advocated in Ref.\cite{z2sf,topth}. 
One feature of this theory is that in the superconducting state $hc/e$ vortices tend to have lower energy than than $hc/2e$ vortices particularly at low doping. 
This is because  in this route to superconductivity  a $hc/2e$ vortex necessarily involves the presence of $Z_2$ gauge flux in its core. 
There is an energy cost to having the $Z_2$ gauge flux which dominates in the low doping limit and raises the energy of the $hc/2e$ vortex. 
 A crucial experimental test\cite{topth} of such a theory is to directly detect 
the `vison' excitation  that carries this conserved $Z_2$ flux, or indirectly to look for signatures of stable $hc/e$ 
vortices\cite{subirhce,nlhce}. Unfortunately, to date, all such experiments have been negative\cite{toexp1,toexp2}. 

This forces us to refine our search further and look for spin liquid states that do not possess $Z_2$ gauge structure ({\em i.e} do not have vison excitations). 
An immediate candidate for such a state is suggested by slave particle mean field
theories of spin-$1/2$ models.
Within a fermionic representation of the spin, a popular mean field state is the $d$-wave
or staggered flux state. This state is conveniently viewed as
a $d$-wave paired state of spinons at $1/2$ filing with the following Hamiltonian:
\begin{equation}
H = -\sum_{<rr'>} \chi_{rr'} (f^{\dagger}_rf_{r'} + h.c) + \Delta_{rr'}(f_{r\uparrow}f_{r'\downarrow} -
(\uparrow \rightarrow \downarrow)).
\end{equation}
Here $f_{r\alpha}$, $\alpha = \uparrow, \downarrow$ is a two-component fermionic spinon
on the site $r$ of a two dimensional square lattice.
The hopping $\chi_{rr'}$ is a real constant, and the pairing $\Delta_{rr'}$
has $d$-wave structure: $\Delta_{rr'} = \Delta_0$ on horizontal bonds and
$\Delta_{rr'} = -\Delta_0$ on vertical bonds. Diagonalization of the mean field
Hamiltonian gives a low energy spectrum with $4$ nodal points
near which a linear dispersing Dirac spectrum appears.

As pointed out in Ref.\cite{affsu2}, this mean field state is equivalent under a
unitary transformation to the `staggered flux' Hamiltonian:
\begin{equation}
H_{sf} = \sum_{<rr'>} \left((i\chi_{rr'} + \Delta_{rr'})f^{\dagger}_rf_{r'} + h.c\right)
\end{equation}
Here we take $r$ to belong to one sublattice of the square lattice so that $r'$ belongs to the opposite sublattice. 
This describes fermionic spin-$1/2$ spinons on the square lattice with complex
hopping amplitudes such that there is a non-zero flux that is staggered from plaquette to
plaquette.

The crucial question is the fate of this mean field state upon including fluctuations.
It is well-known that the important fluctuations involve coupling
to a {\em compact} $U(1)$ gauge field. The resulting Hamiltonian is conveniently written in the 
`staggered flux' gauge and has the structure:
\begin{eqnarray}
H & = & H_f + H_g \\
H_f & = & \sum_{rr'} (i\chi_{rr'} + \Delta_{rr'})e^{ia_{rr'}}f^{\dagger}_rf_{r'} + h.c) \\
H_g & = &  -K \sum_P cos(\vec \nabla \times \vec a) +u\sum_{<rr'>} e_{rr'}^2
\end{eqnarray}
Here $a_{rr'} \in [0,2\pi)$ is to be regarded the spatial component of a $U(1)$ gauge field, and $e_{rr'}$
is the corresponding electric field. Strictly speaking we must take the limit $K \rightarrow 0, u \rightarrow \infty$
but relaxing this condition is not expected to crucially change the physics. The Hamiltonian must be supplemented with a Gauss law 
constraint $\vec \nabla .\vec e + f^{\dagger}_rf_r = 1$ on every lattice site.

Specializing to the low energy limit, this theory may {\em formally} be
viewed as a theory of massless Dirac fermions coupled to a compact $U(1)$
gauge field. Here as usual compactness means that point-like instantons or monopoles are
allowed in the configurations of the gauge field in space-time. At each such monopole
event the gauge flux changes by an integer multiple of $2\pi$. Recent work\cite{stable-u1} has
resolved a long-standing controversy on the low energy behavior of theories of this type,
and shown that
when the number $N$ of Dirac species is large, there is a stable `deconfined' phase
where the instanton fugacity renormalizes to zero at long distances and
low energies. (The case of direct physical interest corresponds to $N = 4$). 
A continuum field theory that describes this low energy fixed point is simply given by the action:
\begin{equation}
\label{u1slet}
S_{fer} = \int d^2x d\tau \bar{\Psi}_\alpha \gamma^\mu (\partial_\mu +i a_\mu) \Psi_\alpha
+ \frac{N}{2 e_g^2} (\epsilon_{\mu\nu\lambda}\partial_\nu a_\lambda)^2 \text{.}
\end{equation}
where $\Psi_{j}$, $j = 1,....,4$ represent the four species of Dirac fermion appropriate for each of the two distinct nodes. 

A precise characterization of the deconfinement\cite{dqcp-science,stable-u1} in this phase is
obtained by noting that the absence of instantons at low energies implies that the gauge flux
$b = \partial_xa_y - \partial_ya_x$ is conserved. This corresponds to an extra global topological $U(1)$ symmetry
at the low energy fixed point that is absent in the microscopic spin model.   

The analysis of Ref. \onlinecite{stable-u1} shows that stable $U(1)$ spin liquids could exist in $d =2$ 
at least for large enough $N$. Whether this stability extends down to 
the physically relevant case of $SU(2)$ spins is not known at present. This corresponds to the case $N = 4$. However the large-$N$ analysis shows that as a matter of principle 
there is no reason to dismiss the possibility of such stable $U(1)$ spin liquids for $SU(2)$ spins. 
For most of the present paper we will simply assume that this is true, and explore its consequences. 
 However
detailed numerical calculation to settle this issue will certainly be useful and welcome. Towards the end of the paper (in Section \ref{alternate})
we will explore the opposite possibility 
- that the stability of the $U(1)$ spin liquids does not extend to $SU(2)$ spins. We show how even in that case it may be possible
to retain the key aspects of the theory provided certain weaker assumptions are satisfied.

This $d$-wave paired $U(1)$ spin liquid state is connected to a conventional
Neel state at ordering vector $(\pi, \pi)$
by a second order transition. This transition will be discussed elsewhere and is an
example of a deconfined quantum critical point\cite{dqcp-science,dqcp-longpaper}. In particular there will be two diverging length scales on the ordered side
- one characterizing the crossover from critical to Neel ordered spin correlations, and a different longer one associated with the 
confinement of spinons. 

A schematic phase diagram depicting the Neel and spin liquid phases at zero
doping is shown in Fig. \ref{sldsc1}.
Now consider doping the Mott insulator. As in many previous works, we assume that
the doping moves the system across the phase boundary
where the magnetism is lost. We may then fruitfully view the doped non-magnetic
state as the result of doping the $U(1)$ spin liquid.

\section{Doped $U(1)$ spin liquids}

What then happens to the $U(1)$ spin liquid when it is doped? 
One answer to this question is suggested by the slave particle mean field theory developed in Ref. \onlinecite{wlsu2} which also provides 
useful mathematical formalism. 
 A $d$-wave superconductor with gapless
nodal quasiparticles is obtained.
 To understand this it is first useful to ask about the
nature of the electric charge carrying excitations in this doped $U(1)$ spin liquid.
Speaking loosely in the presence of fermionic spin-$1/2$
electrically neutral spinons, the charge of the doped holes may be expected to be
associated with spin-$0$, charge-$e$ bosons which carry gauge charge.
This is roughly correct -  more precise consideration shows that there are
two species of such bosons which each carry gauge charge $\pm 1$. These will be
denoted $b_1, b_2$. The presence of two species of
bosons with gauge charges $\pm 1$ (but electric charge $e$ and spin-$0$)
is actually common in situations that involve doping $U(1)$ fractionalized phases.
For instance, in three dimensional fractionalized boson insulators with a
deconfined $U(1)$ gauge field such as that considered in Ref. \onlinecite{MoSeprl02}, two boson
species appear quite naturally. Note that a composite of the two bosons is gauge invariant,
has charge $2e$, and is a spin singlet. Hence it may be identified with a Cooper pair.
Thus the two bosons may each be identified as being one half of a Cooper pair.

Now consider condensing each of the two species of bosons with the same
amplitude $<b_1> = <b_2> \neq 0$. Clearly this state breaks the physical electromagnetic
$U(1)$ symmetry, and hence is superconducting. Further once the bosons condense
the spinons get endowed with electric charge - the structure of the spinon Hamiltonian
imposes $d$-wave symmetry on this superconducting state. In particular the low
energy nodal spinons simply become the nodal quasiparticles of the superconducting state.
This condensate with $<b_1> = <b_2> \neq 0$ is actually energetically favored
in slave particle mean field calculations on the $t-J$ model at low temperature in the low doping
range.

As discussed in the introduction it is instructive to consider the phase diagram not as a function of doping $x$ but as a function of the chemical potential. 
The phase diagram then looks as shown in Fig.\ref{mqc}. As the chemical potential is increased, there is a Mott transition from 
the $U(1)$ spin liquid insulator to the 
$d$-wave superconductor. The corresponding (unstable) zero temperature fixed point controls many aspects of the physics in the underdoped side. In particular 
it determines the universal aspects of the phsyics of the pseudogap regime and its eventual low temperature transition into the superconducting state. 

What is the theory that describes this Mott transition? To understand this we first note that in the insulating state, the two bosons $b_{1,2}$ are gapped. 
Increasing chemical potential decreases this gap. The gap closes at the Mott transition beyond which 
superconductivity is achieved when both boson species condense (with equal amplitude). A continuum theory for the Mott transition that describes 
this condensation is readily written down. It takes the form
\begin{eqnarray}
S & = & S_{fer} + S_{b1} + S_{b2} \\
S_{b1} & = & \int_{\tau,x}  b_1^* \left(\partial_{\tau}- ia_o -\mu - \frac{(\vec \nabla - i\vec a)^2}{2m_b} \right)b_1 \\
S_{b2} & =  & \int_{\tau,x}  b_2^* \left(\partial_{\tau}+ ia_o  - \mu - \frac{(\vec \nabla + i\vec a)^2}{2m_b} \right)b_2
\end{eqnarray}
The change in the sign of the terms involving the gauge potential $(a_0, \vec a)$ reflects the opposite gauge charges carried by $b_1$ and $b_2$. 
The physical hole density is simply the total boson density given by $b_1^*b_1 + b_2^*b_2$, and thus couples linearly to the 
chemical potential $\mu$. 
The fermion part $S_{fer}$ is given in Eqn. \ref{u1slet} above. The transition occurs when $\mu = 0$. As discussed in the introduction a material at fixed doping $x$
will take a path such as that shown in Fig.\ref{mqc} in this phase diagram. 

In writing down this continuum theory we have assumed that the instantons are irrelevant not just at the $U(1)$ spin liquid fixed point but also at the critical
fixed point describing the Mott transition. This is quite reasonable. The presence of extra gapless matter fields is only expected to drive the instantons 
more irrelevant. With this assumption the conservation of gauge flux continues to hold right at this critical fixed point.

In Ref. \onlinecite{wlsu2}, a slave-particle representation with a fermionic charge-$0$ spin-$1/2$ spinon and two species of bosonic charge $e$ spin-$0$ holons
was introduced for the t-J model. Clearly this representation is ideally suited to discuss the zero doping $U(1)$ spin liquid and the corresponding doped system.
In particular the slave particles are `deconfined' in the $U(1)$ spin liquid and the associated Mott critical point to 
the doped superconductor in the sense that instantons in the associated $U(1)$ gauge field are irrelevant.
As emphasized above this leads to an extra global topological symmetry associated with conservation of gauge flux. {\bf This conserved gauge flux
gives precise meaning to the notion of deconfinement and justifies the use of the slave particles as the useful degrees of freedom}.
Indeed the considerations of the previous paragraphs give sharp meaning
to the theory of the underdoped cuprates developed on the basis of this slave particle representation.

\section{Meaning of gauge flux}
As the conserved gauge flux plays a crucial role in this theory it is interesting to ask 
what its meaning is and if its presence can directly be detected
in the non-superconducting state just above $T_c$ in underdoped samples. In the following we describe some physical effects that should obtain if
such a deconfined $U(1)$ gauge field indeed exists. The key is to exploit the unusual structure of the $hc/2e$ vortex that obtains in the superconducting state.
The unusual nature of the vortex stems from the fact that the superconducting state is obtained by condensing separately the two charge-$e$ bosons $b_{1,2}$.
Nevertheless the superconducting state is smoothly connected to a regular BCS $d$-wave superconductor. Thus, despite the condensation of charge-$e$ bosons, the
superconductor must support $hc/2e$ vortices. Such vortex solutions were described by Lee and Wen\cite{lwsu2vrtx}. A simple construction in fact
gives two apparently distinct  $hc/2e$ vortices
- in one there is a full $2\pi$ vortex in $b_1$
but not in $b_2$ and in the other there is a $2\pi$ vortex in $b_2$ but not in $b_1$. Note that the Cooper pair operator $\sim b_1b_2$ winds by $2\pi$ in either case.
To understand these vortex solutions better and as they will prove important for what follows, we consider the following simple energy functional which captures
the basic physics associated with the vortices:
\begin{equation}
\label{sc}
E = \int d^2x \frac{K}{2}\left[ \left(\vec \nabla \theta_1 - \vec a - \vec A \right)^2 + \left(\vec \nabla \theta_2 + \vec a - \vec A \right)^2 \right]
+ .....
\end{equation}
Here $\theta_{1,2}$ are the phases of $b_{1,2}$ respectively. The field $\vec a$ is the vector potential of the internal $U(1)$ gauge field, and $\vec A$ is that
of the
external electromagnetc field.  The ellipses represent various other terms (such as the gauge and external field kinetic energies, etc)
that have not been written out. Let $\theta_{1,2}$ wind by $2\pi m_{1,2}$ on going on a large loop encircling a vortex ($m_{1,2}$ are integers).
Then we must have
\begin{eqnarray}
2\pi m_1 & = & \phi_a + \phi_A \\
2\pi m_2 & = & -\phi_a + \phi_A
\end{eqnarray}
Here $\phi_{a,A}$ are the fluxes of the gauge fields $a, A$ enclosed by the loop. For a $hc/2e$ vortex we must have $\phi_A = \pi$. This will happen if
$m_1 + m_2 = 1$. The internal gauge flux $\phi_a = \pi (m_1 - m_2)$ can then be any {\em odd} integral multiple of $\pi$. In general there wil be some energy cost to
having non-zero internal gauge flux so the lowest value of the gauge flux is energetically preferred. This still leaves the two possibilities $\phi_a = \pm \pi$.
These correspond precisely to the two vortices described above. Ref. \onlinecite{lwsu2vrtx} also pointed out that as the amplitude of only one of the 
two bosons will be suppressed in
the core of either vortex, the region in the core looks like a condensate of $b_1$ but not $b_2$ or vice veras. 
Such condensates can easily be seen to have true staggered
flux order with physical currents circulating in a staggered manner in the core. This staggered current pattern corresponds to a broken symmetry phase
that apparently rears its head in the vortex core. The symmetry of translation by one lattice spacing is broken by the staggering of the current pattern.
However as the core is finite the symmetry cannot truly be broken, and there will be a finite tunneling rate between the two kinds of cores.

It is clear from the discussion above that the two kinds of cores and
the associated staggerd patterns correspond to $\phi_a = \pm \pi$.
We immediately see that the tunneling event between the two kinds of
cores must involve a change of $2\pi$ in the flux of the internal gauge field.
Such a $2\pi$ flux changing event is precisely the instanton in the internal
gauge field. Thus we have the remarkably simple picture of the instanton
as just the tunneling event between the two kinds of staggered flux
cores for $hc/2e$ vortices in the superconducting state.

Note that by assumption the superconducting state descends from a
`normal' state with deconfinement where the instantons are
irrelevant at long scales. However in the superconducting state
the internal gauge flux associated with the vortex is confined to
a flux tube of finite size. The finite size of the flux tube
implies that instantons can no longer necessarily be ignored.
Indeed instantons will also render unstable vortices that have
$\phi_a = \pm 3\pi, \pm 5\pi,.....$ so that there is in principle
a unique $hc/2e$ vortex.

These vortices will presumably start developing integrity in the fluctuation region (see Fig.\ref{mqc}) above the superconducting transition.
For materials with small $T_c$ ({\em i.e} close to the Mott critical point in Fig.), the instanton rate will be small (compared to electronic energy scales). 
Consequently even in the fluctuation region the vortex cores will have staggered flux order that fluctuates slowly at this rate. It is natural to identify this region
with the one observed to have an enhanced Nernst effect in the experiments\cite{xu}. 

The considerations on vortices above enable us to provide some meaning to the gauge flux once there is reasonably well-developed 
local superconducting order ({\em i.e} in the fluctuation region or below). In this region, a useful description is provided by a 
`sigma-model' approach which focuses exclusively on the fluctuating classical order parameters. Indeed such a description was 
obtained in Ref. \onlinecite{leeetalsu2}. Here we will take a slightly different perspective that will bring out 
some key physical aspects. 

The fluctuations of the superconducting order parameter $b^{\dagger}_1b^{\dagger}_2$ are clearly important in this region. As discussed above the vortices in this
order parameter will start having integrity in this region. Since as we have seen above the vortices have slowly fluctuating staggered flux order in their
cores it will be necessary to also include an order parameter for the staggered flux fluctuations. Clearly this is just simply given by 
$n^z \sim |b_1|^2 - |b_2|^2$. Note that both these order parameters (superconducting and staggered flux) have simple descriptions as bilinears in the underlying
bosons.

It is tempting to combine these order parameters into a single three-component vector $\vec n$ with $n^+$ being the pairing order parameter and $n^z$ the staggered
flux one. In the regime where there is well-developed local order one might imagine that this vector has well-defined magnitude but there are fluctuations in 
its direction. A description in terms of a fluctuating unit vector field which represents the local ordering direction may then possibly be appropriate.
The energy functional for this unit vector field will presumably have easy plane anisotropy which favors the superconducting order.  

Despite the appeal of such a description it actually hides a very important piece of physics, and hence is not the full story. The point is that 
in the original description in terms of the boson fields $b_{1,2}$, there is an extra conserved quantity - namely the 
gauge flux -  which should be included in the long wavelength physics. This extra 
conservation arises due to the assumed irrelevance of instantons at the Mott critical fixed point. (Of course at finite temperatures 
close to but not quite at the critical point the instanton fugacity will be small and hence strict conservation of gauge flux will not obtain 
- nevertheless it is necessary to include it to meaningfully discuss physics in the region impacted by the proximity to this critical point). 

What is this conserved flux in terms of the unit vector field $\hat n$? The answer to this is well-known: it is the density  
of defect configurations known as skyrmions. Thus in the absence of instantons, the skyrmion number in the `sigma'-model description is conserved.
It is important to appreciate that this is not a property of familiar quantum $O(3)$ sigma models in two dimensions but is rather closely related to the 
unconventional $O(3)$ models studied in Refs. \onlinecite{laudg,kamrthy,mv}. The latter are models in which hedgehog configurations of the unit vector field have been 
artificially suppressed by hand. In this situation as in the present problem a gauge theoretic description in terms of $CP^1$ fields is superior as it directly 
brings out the extra conservation law. 

An expression for the conserved gauge flux in terms of the observable order parameter fields is readily written down. The gauge magnetic field 
\begin{eqnarray}
{\cal B} & = & 2\pi \rho_{sk} \\
\rho_{sk} & = & \frac{1}{4\pi} \hat{n}. \partial_x \hat{n} \times \partial_y \hat{n}. 
\end{eqnarray}
We write 
\begin{equation}
\hat{n} = (\sin \theta \cos \alpha, \sin \theta \sin \alpha, cos \theta)
\end{equation}
so that $n^z = \cos \theta$ is the staggered flux order parameter, and $\alpha$ is the phase of the superconducting order parameter\footnote{The reader 
should not confuse the
$\vec n$ vector with  vector $\vec I$ introduced in
Ref. ~\protect\onlinecite{leeetalsu2}: 
$\vec I = h^\dagger\vec \sigma h$ with
$h^\dagger = (b_1^*, b_2^*)$ whereas $\vec n = z^\dagger \vec \sigma z$ with $z^\dagger \sim (b_1^*, b_2)$.}
Then the expression above for the gauge magnetic field is readily manipulated to 
\begin{equation}
{\cal B} = \frac{1}{2}\left( \vec \nabla n^z \times \vec \nabla \alpha \right)
\end{equation}
It is easy to check that for a vortex with the structure described above the total gauge flux is $\pm \pi$ as expected.

\section{Detection of gauge flux}
What are some signatures of the gauge flux in experiments? The asymptotic conservation of the gauge flux at the Mott transition fixed point
potentially provides some possibilities for its detection. At non-zero temperatures in the non-superconducting regions, the flux 
conservation is only approximate (as the instanton fugacity is small but non-zero). Nevertheless at low enough temperature the conserved flux will 
propagate diffusively over a long range of length and time scales. Thus there should be an extra diffusive mode that is present at low temperatures in the 
non-superconducting state. It is however not clear how to design a probe that will couple to this diffusive mode at present.

Alternately the vortex structure described above provides a useful way to
create and then detect the gauge flux in the non-superconducting
normal state. We will first describe this by ignoring the
instantons completely in the normal state. The effects of
instantons will then be discussed.

\begin{figure}[tb]
\centering
\includegraphics[width=3.0in]{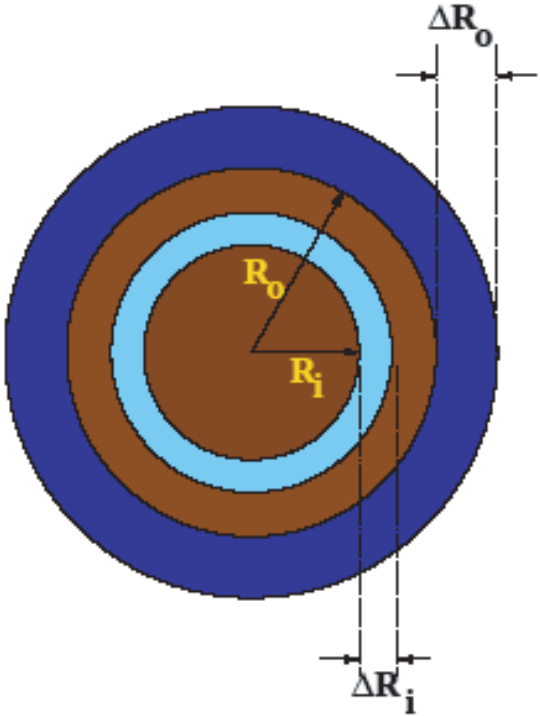}
\caption{Structure of the sample needed for the proposed experiment. The outer annulus (in dark blue) has the highest $T_c$. The inner annulus 
(in light blue) has a smaller $T_c$. The rest of the sample (in brown) has even smaller $T_c$. 
 } \label{u1ft1}
\end{figure}

Consider first a large disc of
cuprate material which is such that the doping level changes as a function of the radial distance from the center as shown in Fig. \ref{u1ft1}.
The outermost annulus has the largest doping $x_1$. The inner annulus has a
lower doping level $x_2$. The rest of the sample is at a doping level
$x_3< x_2 < x_1$. The corresponding transition temperatures $T_{c1,2,3}$
will be such that $T_{c3} < T_{c2} < T_c{1}$. We also imagine that
the thickness $\Delta R_o, \Delta R_i$ of the outer and inner  annuli
are both much smaller than the penetration
depth for the physical vector potential $A$. The penetration depth
of the internal gauge field $a$ is expected to be small and we
expect it will be smaller than $\Delta R_o, \Delta R_i$.  
We also imagine that the radius of this inner annulus $R_i$ is a substantial fraction of the radius $R_o$
of the outer annulus.

\begin{figure}[tb]
\centering
\includegraphics[width=3.0in]{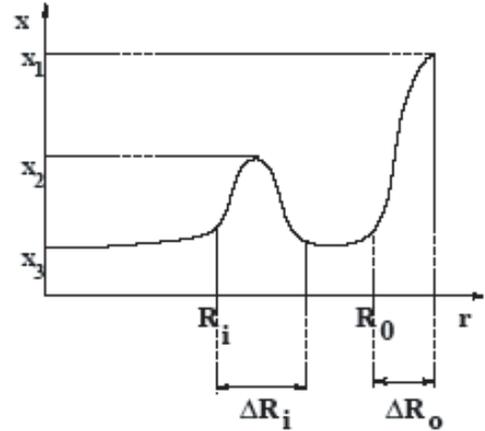}
\caption{Variation of the doping level $x$ with radial distance $r$ in the sample. 
 } \label{u1ft2}
\end{figure}

Now consider the following set of operations on such a sample.

(i) First cool in a magnetic field to a temperature $T_{in}$ such that
$T_{c2} < T_{in} < T_{c1}$. The outer ring will then go superconducting while the rest of the sample stays normal.
In the presence of the field the outer ring will condense into a state in which there is a net vorticity on going around the ring.
We will be interested in the case where this net vorticity is an odd multiple of the basic $hc/2e$ vortex. If as assumed the
physical penetration depth is much bigger than the thickness $\Delta R_o$ then the physical magnetic flux enclosed by the ring will not be quantized.

(ii) Now consider turning off
the external magetic field. The vortex present in the outer superconducting ring will stay (manifested as a small circulating persistent current)
and will give rise to a small magnetic field. As explained above if the vorticity is odd, then it must be associated with a flux of the
internal gauge field that is $\pm \pi$. This internal gauge flux must essentially all be in the inner `normal' region of the sample with
very small penetration into the outer superconducting ring. It will spread out essentially evenly over the full inner region.
We have thus managed to create a configuration with a non-zero  internal gauge flux
in the non-superconducting state.

(iii) How do we detect the presence of this internal gauge flux? For that imagine now cooling the sample further to a temperature
$T_{fin}$ such that $T_{c3} < T_{fin} < T_{c2}$. Then the inner ring will also go superconducting. This is to be understood as the condensation
of the two boson species $b_{1,2}$. But this condensation occurs in the presence of some internal gauge flux. When the bosons $b_{1,2}$ condense in the inner ring,
they will do so in a manner that quantizes the internal gauge flux enclosed by this inner ring into an integer multiple of $\pi$. If as
assumed the inner radius is a substantial fraction of the outer radius then the net internal gauge flux will prefer the quantized values $\pm \pi$ rather than
be zero (see below). However configurations of the inner ring that enclose quantized internal gauge flux of $\pm \pi$ also necessarily
contain a physical vortex that is an odd multiple of $hc/2e$. With the thickness of the inner ring being smaller than the physical penetration depth, 
most of the physical magnetic flux will escape. There will still be a small residual physical flux due to the current in the inner ring associated with the 
induced vortex. This residual physical magnetic flux can then be detected.

Note that the sign of the induced physical flux is independent of the sign of the initial magnetic field. Furthermore
the effect obtains only if the initial vorticity in the outer ring is odd. If on the other hand the initial vorticity is
even the associated internal gauge flux is zero, and there will be no induced physical flux when the inner ring goes superconducting.

The preceding discussion ignores any effects of instantons. In
contrast to a bulk vortex in the superconducting state the
vortices in the set-up above have macroscopic cores. The internal
gauge flux is therefore distributed over a region of macroscopic
size. Consequently if instantons are irrelevant at long scales in
the normal state, their rate may be expected to be small. At any
non-zero temperature (as in the proposed experiment) there will be
a non-zero instanton rate which will be small for small
temperature.

When such instantons are allowed then the internal gauge flux
created in the sample after step (ii) will fluctuate between the
values $+\pi$ and $-\pi$. However so long as the time required to
form the physical vortex in step (iii) is much longer than this
instanton rate we expect that the effect will be seen.

\subsection{Energy estimates}
In this subsection we take a closer look at the energies involved with various conceivable outcomes of the process described above.
We assume that the cooling is slow enough that the system always stays in equillibrium so that the outcome of the experiment is determined by 
thermodynamic considerations.  
This will give us some indication of the conditions under which the experiment has its best chance of giving a positive 
result. We will use the simple energy functional in Eqn. \ref{sc} but will now need to keep track of the kinetic energy terms for the 
internal gauge field. In principle we should also account for the existence of many weakly coupled layers in the three dimensional sample and 
the energy of the external magnetic field. To begin with we will ignore these latter effects and comment on them later. 

The energy per layer in the superconducting state is thus given by 
\begin{eqnarray}
E & = & E_b + E_a \\
E_b & = & \int d^2x \frac{K}{2}\left[ \left(\vec \nabla \theta_1 - \vec a - \vec A \right)^2 + \left(\vec \nabla \theta_2 + \vec a - \vec A \right)^2 \right] \\
E_a & = & \int d^2x g \left(\vec \nabla \times \vec a \right)^2
\end{eqnarray}
Here $K$ is the superfluid stiffness and is roughly of the order of $k_BT_c$. The parameter $g$ is more difficult to estimate 
in any reliable manner. A crude estimate is provided by the following consideration. Right at the Mott critical point, we may 
hope to calculate the energy associated with various configurations of the gauge field in a `RPA' approach which integrates out both the spinon and boson fields. 
In this calculation the gauge action is dominated by contributions from the spinons and is roughly given by 
\begin{equation}
S_{eff}[a] \sim Ja\int d^2q d\omega \sqrt{\omega^2 + q^2}|\vec a ^T(\vec q, \omega)|^2
\end{equation} 
with $J$ the exchange energy and $a$ the lattice spacing (not to be confused with the vector potential). For the energy, this corresponds to an effective  
coupling $g$ that is singular: $g(q) \sim Ja/q$ for small $q$. At finite temperature $T$, this singular behaviour will be cutoff at a value 
$q_T \sim k_B T/v_F \sim k_B T/Ja$. We thus have the estimate
\begin{equation}
g \approx \frac{J^2 a^2}{k_B T}.
\end{equation}
We caution however that this estimate is very crude. 

The energy associated with a supercurrent (the $K$ term) competes with the energy associated with the internal gauge field $\vec a$. 
At the end of the experiment discussed above, the system has one of two options. It can choose to quantize the gauge flux in the inner ring at $\pi$
while paying the cost of a spontaneous supercurrent through the inner annulus. The other option is that it prferes to push all the 
internal gauge flux out into the region between the two superconducting annuli thereby creating no supercurrent. If this were to occur the experiment will give
a null result. It is therefore necessary to compare the energies of these two possible outcomes. 

The energy cost of the supercurrent is readily estimated and is 
\begin{equation}
E_J \sim (\Delta R_i) R_i \left(\frac{K}{R_i^2}\right) \sim T_c \left(\frac{\Delta R_i}{R_i} \right)
\end{equation}
For a sample with $T_c \sim 10 K$, and $\Delta R_i \sim 0.1 R_i$, we have $E_J \sim 1 K$. 
The energy cost of pushing the field out of the inner ring is roughly
\begin{equation}
E_a \sim g\frac{1}{R_o^2 - R_i^2} \sim \frac{J^2a^2}{k_B T R_o^2f}
\end{equation}
where $f\pi R_o^2$ is the area between the inner and outer rings. If we assume $f \sim 0.1$, then 
$E_a > 1 K$ for samples with $R_o < \approx 1000 a$. Thus the energy cost per layer of the gauge field may be expected to 
dominate for samples of outer diameter about or smaller than a micron. 
However in view of the crudeness of the energy estimates, this should only be taken as
very rough guidance. 

It is clearly also advantageous to have several layers so that the total energy gain in forming the supercurrent is
multiplied by the number of layers and can exceed $k_B T$. Finally we also mention that if the inner annulus thickness is much bigger than the 
physical penetration depth then all of the physical flux associated with the trapped inner vortex will stay in the sample. 
However the energy cost of this physical flux is easily seen to be enormous - it overwhelmes any effects due to the internal gauge field. 
It is for this reason that we advocate using an inner annulus thinner than the physical penetration depth to let most of the flux escape.

\section{Alternate theoretical possibilies}
\label{alternate}
In this Section, we wish to sketch some alternate theoretical possibilities that weaken the assumptions made in previous Sections. 
In particular let us contemplate the situation that the stability of the $U(1)$ spin liquid state found in 
the limit of a large number $N$ of species fo Dirac spinons does not extend to the physically relevant case of $N = 4$. 
In other words, what if such a $U(1)$ spin liquid is never a stable phase for any model of $SU(2)$ spins with short ranged interactions in two dimensions? 
If the instability is due to the relevance of a monopole operator then it is natural to expect confinement to occur. A natural possible confined state 
that can result is simply the conventional Neel antiferromagnet\footnote{While this is certainly a reasonable expectation, the Neel state 
has never been satisfactorily described this way. Work on this is in progress\cite{hermunpub}}.

We now show how even in this case it may still be possible to view the underdoped cuprates as doped $U(1)$ spin liquids for the physics in 
an intermediate window of temperatures. This window will describe the pseudogap regime. Our arguments will rely 
crucially on lessons drawn from recent work\cite{dqcp-science,dqcp-longpaper} on deconfined quantum criticality. Indeed Ref. \onlinecite{dqcp-science,dqcp-longpaper} 
showed that {\em unstable gapped} $U(1)$ spin liquids may emerge as good descriptions of the physics over a broad intermediate region 
of length and time scales close to certain quantum phase transitions. Here we will argue that in a similar vein the $U(1)$ spin liquid state with gapless 
Dirac spinons considered in this paper may also control the physics in a broad intermediate regime near certain quantum transitions at zero doping even 
if it is eventually unstable toward confinement. We will then discuss the implications for the doped system. 

We now consider the possibility that the spin liquid state in Figs. \ref{sldsc1}, \ref{sldsc2} is  
a $Z_2$ state of the spin system that has nodal Dirac spinons. We will argue that a large region near the transition between this state and  the 
conventional  Neel state may be controlled by the (unstable) $U(1)$ spin liquid fixed point. We recall that the $Z_2$ 
state also has a gapped `vison' excitation that carries the $Z_2$ 
gauge flux. Now consider a confinement transition out of this state that is obtained by condensing the vison. For {\em gapped} $Z_2$ spin liquids, such 
a transition to a confined valence bond solid state is possible. 
We refer the reader to Ref. \onlinecite{dqcp-longpaper} for a complete discussion. As discussed there, 
this transition provides an example of a deconfined quantum critical point. In particular
the critical theory is that of a critical charged boson (interpreted as a spinon pair field) 
coupled to a fluctuating non-compact $U(1)$ gauge field.

To study the confinement transition out of the $Z_2$ spin liquid with gapless nodal spinons, we follow the same general idea as in Ref. \onlinecite{dqcp-longpaper}.
Such a state is conveniently described as a theory of Dirac spinons coupled to a {\em compact} $U(1)$ gauge field provided we also condense a singlet pair of 
spinons. Such a spinon pair has gauge charge-$2$ so that when it condenses, the $U(1)$ gauge theory enters a Higgs phase whose universal 
physics is described in terms of a deconfined $Z_2$ gauge theory. Schematically consider an action with the following general structure
\begin{eqnarray}\label{z2afm}
S & = & S_{fer} + S_{f\phi} + S_{sp} \\
S_{f\phi} & = & \int d^2x d\tau \kappa \left(\phi_{sp} \psi^T \sigma^y \psi + h.c\right) \\
S_{sp} & = & \int d^2x d\tau |(\partial_{\mu} - 2ia_{\mu})\phi_{sp}|^2 + ....
\end{eqnarray}
Here $\phi_{sp}$ represents the spinon pair field - this is taken to be invariant under all the global symmetries of the microscopic spin model. 
As mentioned above the $Z_2$ spin liquid appears as the Higgs phase where $\phi_{sp}$ is condensed. 
Now consider a phase transition where $\phi_{sp}$
uncondenses, {\em i.e} gets gapped. To discuss the low energy physics of such a phase we simply drop terms involving $\phi_{sp}$. The result is precisely 
the theory of Dirac spinons coupled to a compact $U(1)$ gauge field. In this Section we are assuming that monopoles are relevant at the 
$U(1)$ spin liquid fixed point for the physical case $N = 4$ ($N$ is the number of Dirac species). The resulting state is expected to be confined, presumably just the usual 
Neel state. 

It is however perfectly possible that monopole events though relevant at the $U(1)$ spin liquid fixed point are irrelevant at the critical fixed point 
controlling the transition to the Higgs phase ({\em i.e} the $Z_2$ spin liquid). Indeed this is precisely what happens in the situations with gapped spinons 
considered in Ref. \onlinecite{dqcp-science,dqcp-longpaper}. A particularly close analog is provided by the transition between a valence bond solid paramagnet and a 
spin gapped $Z_2$ paramagnet. Furthermore in the present gapless case, the critical fixed point has a greater number of gapless matter fields than in the $U(1)$ 
spin liquid fixed point - increasing the number of gapless matter fields is expected to increase the scaling dimension of the monopoles. Thus it appears possible 
that monopoles are irrelevant at the critical fixed point though relevant at the $U(1)$ spin liquid fixed point. In the following we will assume that this is the case. 
This is a weaker assumption than the one made in previous sections.

\begin{figure}[tb]
\centering
\includegraphics[width=3.0in]{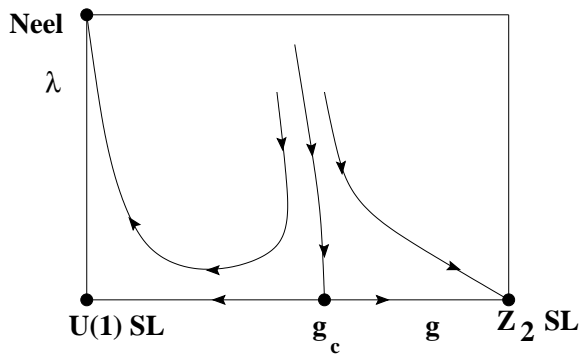}
\caption{Possible renormalization group flow diagram that allows for a direct seconf order transition between a conventional collinear Neel magnet, and a 
$Z_2$ spin liquid with gapless Dirac spinons. $g$ is the parameter used to tune through the transition. $\lambda$ is the fugacity of the leading allowed monopole operator. 
The $U(1)$ spin liquid with gapless Dirac spinons appears as an unstable fixed point that nevertheless controls the physics in a broad intermediate window 
of length scales in the ordered side of the phase transition.  
 } \label{rgz2afm}
\end{figure}

\begin{figure}[tb]
\centering
\includegraphics[width=3.0in]{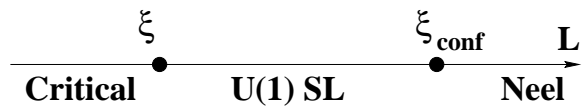}
\caption{Crossovers as a function of length scale in the Neel state close to the transition assuming the validity of the RG flows of Fig. \protect{\ref{rgz2afm}}. 
There is a broad intermediate regime where the physics is that of gapless Dirac spinons in a $U(1)$ spin liquid state. 
 } \label{afmls}
\end{figure}

Within this weaker assumption, a number of interesting possibilities arise. First it strongly suggests the 
possibility of a direct second order transition between a $Z_2$ spin liquid with Dirac spinons and a conventional collinear Neel 
antiferromagnet. The critical point is described by the theory in Eqns. \ref{z2afm} and where the compactness may be ignored at low energies. This is 
a deconfined quantum critical point in the same sense as in Ref. \onlinecite{dqcp-science,dqcp-longpaper}. In particular the flux of the $U(1)$ gauge field is conserved
at the low energy fixed point. The assumed renormalization group flows near this critical fixed point are shown in Fig. \ref{rgz2afm}. As in other situations 
with these flows, on approaching the transition from the Neel side, there will be two diverging length scales $\xi$ and $\xi_{conf}$ with $\xi_{conf}$ diverging 
as a power of $\xi$. Thus $\xi_{conf}$ is potentially much bigger than $\xi$ near the transition. The first length scale $\xi$ is the scale below which 
the correlations appear to be that of the critical fixed point. At length scales bigger then $\xi$ but smaller than $\xi_{conf}$, a description in terms of the 
$U(1)$ spin liquid with Dirac spinons is appropriate. Finally it is only on length scales bigger than $\xi_{conf}$ does the system crossover to the
confined Neel state. Thus though unstable the $U(1)$ spin liquid would (within the assumptions above) still describe the physics in a broad intermediate region 
near the transition between the  Neel state and a $Z_2$ spin liquid with Dirac spinons.

What may we then say about the doped system? It is now only necessary to assume that doping effectively pushes the system {\em toward} the critical point 
to the $Z_2$ spin liquid. Then it is natural to expect that the doped version of the $U(1)$ spin liquid provides a description of the physics in a broad intermediate
regime of length and energy scales. This intermediate regime may possibly describe the pseudogap region in moderately doped samples. 

At present we do not know which (if either) of the two possibilities for the fate of the $U(1)$ spin liquid discussed above obtains. 
We hope that the considerations here will set the stage for future work on these questions.

\section{Conclusion}
We have pursued the old idea that the underdoped metallic cuprates are fruitfully viewed as doped non-magnetic Mott insulators. We sharpened the 
meaning of this notion by first considering the phase diagram as a function of chemical potential rather than hole density. We suggested that 
the doped metal may be understood as being close to a chemical potential tuned quantum transition from a spin liquid Mott insulator 
to the $d$-wave superconductor. From this perspective the actual detailed path followed by the real material as it evolves from the undoped antiferromagnetic 
Mott insulator to the doped metal is not that important in determining many aspects of the latter. Indeed this path in the real material appears
to be complicated and probably non-universal (in that it varies between different cuprates). What is universal is the behavior in a wide window of intermediate temperatures 
(or energy scale) in the doped metal. Our suggestion is that this physics is captured by the universal properties of the Mott critical point between the spin liquid insulator 
and the $d$-wave superconductor. 

The particular spin liquid state we considered has gapless nodal spinons coupled to a fluctuating non-compact $U(1)$ gauge field. These
spinons may be considered `deconfined'. We emphasize that this term does {\em not} mean that the spinons behave as free fermions because they are still 
strongly coupled to 
the fluctuating $U(1)$ gauge field\cite{RWspin}. Rather the precise meaning to the notion of deconfinement 
is provided by the observation that there is an extra emergent global (topological) conservation 
law in this
phase that is not present at the microscopic level. This conserved quantity is simply the gauge flux. We discussed the meaning of the gauge flux, and suggested 
physical effects that could help detect its existence in experiments.

\begin{acknowledgments}
We thank Kathryn A. Moler, Mike Hermele, and Matthew P.A. Fisher for useful discussions. 
This research is supported by the
National Science Foundation grant DMR-0308945 (T.S.), DMR--02--01069 (PAL).  T.S also
acknowledges funding from the NEC Corporation, the Alfred P. Sloan Foundation,
and an award from The Research Corporation.

\end{acknowledgments}

\vspace{0.5in}

\bibliography{u1ft}

\begin{thebibliography}{27}
\expandafter\ifx\csname natexlab\endcsname\relax\def\natexlab#1{#1}\fi
\expandafter\ifx\csname bibnamefont\endcsname\relax
  \def\bibnamefont#1{#1}\fi
\expandafter\ifx\csname bibfnamefont\endcsname\relax
  \def\bibfnamefont#1{#1}\fi
\expandafter\ifx\csname citenamefont\endcsname\relax
  \def\citenamefont#1{#1}\fi
\expandafter\ifx\csname url\endcsname\relax
  \def\url#1{\texttt{#1}}\fi
\expandafter\ifx\csname urlprefix\endcsname\relax\def\urlprefix{URL }\fi
\providecommand{\bibinfo}[2]{#2}
\providecommand{\eprint}[2][]{\url{#2}}

\bibitem[{\citenamefont{Orenstein and Millis}(2000)}]{ormil}
\bibinfo{author}{\bibfnamefont{J.}~\bibnamefont{Orenstein}} \bibnamefont{and}
  \bibinfo{author}{\bibfnamefont{A.~J.} \bibnamefont{Millis}},
  \bibinfo{journal}{Science} \textbf{\bibinfo{volume}{288}},
  \bibinfo{pages}{468} (\bibinfo{year}{2000}).

\bibitem[{\citenamefont{Anderson}(1987)}]{pwa}
\bibinfo{author}{\bibfnamefont{P.~W.} \bibnamefont{Anderson}},
  \bibinfo{journal}{Science} \textbf{\bibinfo{volume}{235}},
  \bibinfo{pages}{1196} (\bibinfo{year}{1987}).

\bibitem[{\citenamefont{Sachdev}(2003)}]{subirrmp}
\bibinfo{author}{\bibfnamefont{S.}~\bibnamefont{Sachdev}},
  \bibinfo{journal}{Rev. Mod. Phys} \textbf{\bibinfo{volume}{75}},
  \bibinfo{pages}{913} (\bibinfo{year}{2003}).

\bibitem[{\citenamefont{Affleck and Marston}(1988)}]{affleck-marston-rapid}
\bibinfo{author}{\bibfnamefont{I.}~\bibnamefont{Affleck}} \bibnamefont{and}
  \bibinfo{author}{\bibfnamefont{J.~B.} \bibnamefont{Marston}},
  \bibinfo{journal}{Phys.\ Rev.\ B} \textbf{\bibinfo{volume}{37}},
  \bibinfo{pages}{3774} (\bibinfo{year}{1988}).

\bibitem[{\citenamefont{Wen and Lee}(1996)}]{wlsu2}
\bibinfo{author}{\bibfnamefont{X.-G.} \bibnamefont{Wen}} \bibnamefont{and}
  \bibinfo{author}{\bibfnamefont{P.~A.} \bibnamefont{Lee}},
  \bibinfo{journal}{Phys. Rev. Lett.} \textbf{\bibinfo{volume}{76}},
  \bibinfo{pages}{503} (\bibinfo{year}{1996}).

\bibitem[{\citenamefont{Kim and Lee}(1999)}]{KL9930}
\bibinfo{author}{\bibfnamefont{D.~H.} \bibnamefont{Kim}} \bibnamefont{and}
  \bibinfo{author}{\bibfnamefont{P.~A.} \bibnamefont{Lee}},
  \bibinfo{journal}{Annals of Physics} \textbf{\bibinfo{volume}{272}},
  \bibinfo{pages}{130} (\bibinfo{year}{1999}).

\bibitem[{\citenamefont{Rantner and Wen}(2002)}]{RWspin}
\bibinfo{author}{\bibfnamefont{W.}~\bibnamefont{Rantner}} \bibnamefont{and}
  \bibinfo{author}{\bibfnamefont{X.-G.} \bibnamefont{Wen}},
  \bibinfo{journal}{Phys. Rev. B} \textbf{\bibinfo{volume}{66}},
  \bibinfo{pages}{144501} (\bibinfo{year}{2002}).

\bibitem[{\citenamefont{Hermele et~al.}({\natexlab{a}})\citenamefont{Hermele,
  Senthil, Fisher, Lee, Nagaosa, and Wen}}]{stable-u1}
\bibinfo{author}{\bibfnamefont{M.}~\bibnamefont{Hermele}},
  \bibinfo{author}{\bibfnamefont{T.}~\bibnamefont{Senthil}},
  \bibinfo{author}{\bibfnamefont{M.~P.~A.} \bibnamefont{Fisher}},
  \bibinfo{author}{\bibfnamefont{P.~A.} \bibnamefont{Lee}},
  \bibinfo{author}{\bibfnamefont{N.}~\bibnamefont{Nagaosa}}, \bibnamefont{and}
  \bibinfo{author}{\bibfnamefont{X.-G.} \bibnamefont{Wen}},
  \eprint{cond-mat/0404751}.

\bibitem[{\citenamefont{Read and Sachdev}(1989)}]{ReSaSuN}
\bibinfo{author}{\bibfnamefont{N.}~\bibnamefont{Read}} \bibnamefont{and}
  \bibinfo{author}{\bibfnamefont{S.}~\bibnamefont{Sachdev}},
  \bibinfo{journal}{Phys.\ Rev.\ Lett.} \textbf{\bibinfo{volume}{62}},
  \bibinfo{pages}{1694} (\bibinfo{year}{1989}).

\bibitem[{\citenamefont{Vojta and Sachdev}(1999)}]{SaVoj}
\bibinfo{author}{\bibfnamefont{M.}~\bibnamefont{Vojta}} \bibnamefont{and}
  \bibinfo{author}{\bibfnamefont{S.}~\bibnamefont{Sachdev}},
  \bibinfo{journal}{Phys.\ Rev.\ Lett.} \textbf{\bibinfo{volume}{83}},
  \bibinfo{pages}{396} (\bibinfo{year}{1999}).

\bibitem[{\citenamefont{Senthil and Fisher}(2000)}]{z2sf}
\bibinfo{author}{\bibfnamefont{T.}~\bibnamefont{Senthil}} \bibnamefont{and}
  \bibinfo{author}{\bibfnamefont{M.~P.~A.} \bibnamefont{Fisher}},
  \bibinfo{journal}{Phys. Rev. B} \textbf{\bibinfo{volume}{62}},
  \bibinfo{pages}{7850} (\bibinfo{year}{2000}).

\bibitem[{\citenamefont{Senthil and Fisher}(2001)}]{topth}
\bibinfo{author}{\bibfnamefont{T.}~\bibnamefont{Senthil}} \bibnamefont{and}
  \bibinfo{author}{\bibfnamefont{M.~P.~A.} \bibnamefont{Fisher}},
  \bibinfo{journal}{Phys. Rev. Lett.} \textbf{\bibinfo{volume}{86}},
  \bibinfo{pages}{292} (\bibinfo{year}{2001}).

\bibitem[{\citenamefont{Sachdev}(1992)}]{subirhce}
\bibinfo{author}{\bibfnamefont{S.}~\bibnamefont{Sachdev}},
  \bibinfo{journal}{Phys. Rev. B} \textbf{\bibinfo{volume}{45}},
  \bibinfo{pages}{389} (\bibinfo{year}{1992}).

\bibitem[{\citenamefont{Nagaosa and Lee}(1992)}]{nlhce}
\bibinfo{author}{\bibfnamefont{N.}~\bibnamefont{Nagaosa}} \bibnamefont{and}
  \bibinfo{author}{\bibfnamefont{P.~A.} \bibnamefont{Lee}},
  \bibinfo{journal}{Phys. Rev. B} \textbf{\bibinfo{volume}{45}},
  \bibinfo{pages}{966} (\bibinfo{year}{1992}).

\bibitem[{\citenamefont{Wynn et~al.}(2001)\citenamefont{Wynn, Bonn, Gardner,
  Lin, Liang, Hardy, Kirtley, and Moler}}]{toexp1}
\bibinfo{author}{\bibfnamefont{J.}~\bibnamefont{Wynn}},
  \bibinfo{author}{\bibfnamefont{D.~A.} \bibnamefont{Bonn}},
  \bibinfo{author}{\bibfnamefont{B.~W.} \bibnamefont{Gardner}},
  \bibinfo{author}{\bibfnamefont{Y.-J.} \bibnamefont{Lin}},
  \bibinfo{author}{\bibfnamefont{R.}~\bibnamefont{Liang}},
  \bibinfo{author}{\bibfnamefont{W.~N.} \bibnamefont{Hardy}},
  \bibinfo{author}{\bibfnamefont{J.~R.} \bibnamefont{Kirtley}},
  \bibnamefont{and} \bibinfo{author}{\bibfnamefont{K.~A.} \bibnamefont{Moler}},
  \bibinfo{journal}{Phys. Rev. Lett.} \textbf{\bibinfo{volume}{87}},
  \bibinfo{pages}{197002} (\bibinfo{year}{2001}).

\bibitem[{\citenamefont{Bonn et~al.}(2001)\citenamefont{Bonn, Wynn, Gardner,
  Liang, Hardy, Kirtley, and Moler}}]{toexp2}
\bibinfo{author}{\bibfnamefont{D.~A.} \bibnamefont{Bonn}},
  \bibinfo{author}{\bibfnamefont{J.~C.} \bibnamefont{Wynn}},
  \bibinfo{author}{\bibfnamefont{B.~W.} \bibnamefont{Gardner}},
  \bibinfo{author}{\bibfnamefont{R.}~\bibnamefont{Liang}},
  \bibinfo{author}{\bibfnamefont{W.~N.} \bibnamefont{Hardy}},
  \bibinfo{author}{\bibfnamefont{J.~R.} \bibnamefont{Kirtley}},
  \bibnamefont{and} \bibinfo{author}{\bibfnamefont{K.~A.} \bibnamefont{Moler}},
  \bibinfo{journal}{Nature} \textbf{\bibinfo{volume}{414}},
  \bibinfo{pages}{887} (\bibinfo{year}{2001}).

\bibitem[{\citenamefont{Affleck et~al.}(1988)\citenamefont{Affleck, Zhou, Hsu,
  and Anderson}}]{affsu2}
\bibinfo{author}{\bibfnamefont{I.}~\bibnamefont{Affleck}},
  \bibinfo{author}{\bibfnamefont{Z.}~\bibnamefont{Zhou}},
  \bibinfo{author}{\bibfnamefont{T.}~\bibnamefont{Hsu}}, \bibnamefont{and}
  \bibinfo{author}{\bibfnamefont{P.}~\bibnamefont{Anderson}},
  \bibinfo{journal}{Phys. Rev. B} \textbf{\bibinfo{volume}{38}},
  \bibinfo{pages}{745} (\bibinfo{year}{1988}).

\bibitem[{\citenamefont{Senthil et~al.}(2004)\citenamefont{Senthil, Vishwanath,
  Balents, Sachdev, and Fisher}}]{dqcp-science}
\bibinfo{author}{\bibfnamefont{T.}~\bibnamefont{Senthil}},
  \bibinfo{author}{\bibfnamefont{A.}~\bibnamefont{Vishwanath}},
  \bibinfo{author}{\bibfnamefont{L.}~\bibnamefont{Balents}},
  \bibinfo{author}{\bibfnamefont{S.}~\bibnamefont{Sachdev}}, \bibnamefont{and}
  \bibinfo{author}{\bibfnamefont{M.~P.~A.} \bibnamefont{Fisher}},
  \bibinfo{journal}{Science} \textbf{\bibinfo{volume}{303}},
  \bibinfo{pages}{1490} (\bibinfo{year}{2004}).

\bibitem[{\citenamefont{Senthil et~al.}()\citenamefont{Senthil, Balents,
  Sachdev, Vishwanath, and Fisher}}]{dqcp-longpaper}
\bibinfo{author}{\bibfnamefont{T.}~\bibnamefont{Senthil}},
  \bibinfo{author}{\bibfnamefont{L.}~\bibnamefont{Balents}},
  \bibinfo{author}{\bibfnamefont{S.}~\bibnamefont{Sachdev}},
  \bibinfo{author}{\bibfnamefont{A.}~\bibnamefont{Vishwanath}},
  \bibnamefont{and} \bibinfo{author}{\bibfnamefont{M.~P.~A.}
  \bibnamefont{Fisher}}, \eprint{cond-mat/0312617}.

\bibitem[{\citenamefont{Motrunich and Senthil}(2002)}]{MoSeprl02}
\bibinfo{author}{\bibfnamefont{O.}~\bibnamefont{Motrunich}} \bibnamefont{and}
  \bibinfo{author}{\bibfnamefont{T.}~\bibnamefont{Senthil}},
  \bibinfo{journal}{Phys. Rev. Lett.} \textbf{\bibinfo{volume}{89}},
  \bibinfo{pages}{277004} (\bibinfo{year}{2002}).

\bibitem[{\citenamefont{Lee and Wen}(2001)}]{lwsu2vrtx}
\bibinfo{author}{\bibfnamefont{P.~A.} \bibnamefont{Lee}} \bibnamefont{and}
  \bibinfo{author}{\bibfnamefont{X.-G.} \bibnamefont{Wen}},
  \bibinfo{journal}{Phys. \ Rev. \ B} \textbf{\bibinfo{volume}{63}},
  \bibinfo{pages}{224517} (\bibinfo{year}{2001}).

\bibitem[{\citenamefont{Xu et~al.}(2000)\citenamefont{Xu, Ong, Wang, Kakeshita,
  and Uchida}}]{xu}
\bibinfo{author}{\bibfnamefont{Z.~A.} \bibnamefont{Xu}},
  \bibinfo{author}{\bibfnamefont{N.~P.} \bibnamefont{Ong}},
  \bibinfo{author}{\bibfnamefont{Y.}~\bibnamefont{Wang}},
  \bibinfo{author}{\bibfnamefont{T.}~\bibnamefont{Kakeshita}},
  \bibnamefont{and} \bibinfo{author}{\bibfnamefont{S.}~\bibnamefont{Uchida}},
  \bibinfo{journal}{Nature} \textbf{\bibinfo{volume}{406}},
  \bibinfo{pages}{486} (\bibinfo{year}{2000}).

\bibitem[{\citenamefont{Lee et~al.}(1998)\citenamefont{Lee, Nagaosa, Ng, and
  Wen}}]{leeetalsu2}
\bibinfo{author}{\bibfnamefont{P.~A.} \bibnamefont{Lee}},
  \bibinfo{author}{\bibfnamefont{N.}~\bibnamefont{Nagaosa}},
  \bibinfo{author}{\bibfnamefont{T.-K.} \bibnamefont{Ng}}, \bibnamefont{and}
  \bibinfo{author}{\bibfnamefont{X.-G.} \bibnamefont{Wen}},
  \bibinfo{journal}{Phys. Rev. B} \textbf{\bibinfo{volume}{57}},
  \bibinfo{pages}{6003} (\bibinfo{year}{1998}).

\bibitem[{\citenamefont{Lau and Dasgupta}(1989)}]{laudg}
\bibinfo{author}{\bibfnamefont{M.}~\bibnamefont{Lau}} \bibnamefont{and}
  \bibinfo{author}{\bibfnamefont{C.}~\bibnamefont{Dasgupta}},
  \bibinfo{journal}{Phys. Rev. B} \textbf{\bibinfo{volume}{39}},
  \bibinfo{pages}{7212} (\bibinfo{year}{1989}).

\bibitem[{\citenamefont{Kamal and Murthy}(1993)}]{kamrthy}
\bibinfo{author}{\bibfnamefont{M.}~\bibnamefont{Kamal}} \bibnamefont{and}
  \bibinfo{author}{\bibfnamefont{G.}~\bibnamefont{Murthy}},
  \bibinfo{journal}{Phys. Rev. Lett.} \textbf{\bibinfo{volume}{71}},
  \bibinfo{pages}{1911} (\bibinfo{year}{1993}).

\bibitem[{\citenamefont{Motrunich and Vishwanath}()}]{mv}
\bibinfo{author}{\bibfnamefont{O.}~\bibnamefont{Motrunich}} \bibnamefont{and}
  \bibinfo{author}{\bibfnamefont{A.}~\bibnamefont{Vishwanath}},
  \eprint{cond-mat/0311222}.

\bibitem[{\citenamefont{Hermele et~al.}({\natexlab{b}})\citenamefont{Hermele,
  Senthil, and Fisher}}]{hermunpub}
\bibinfo{author}{\bibfnamefont{M.}~\bibnamefont{Hermele}},
  \bibinfo{author}{\bibfnamefont{T.}~\bibnamefont{Senthil}}, \bibnamefont{and}
  \bibinfo{author}{\bibfnamefont{M.~P.~A.} \bibnamefont{Fisher}},
  \emph{\bibinfo{title}{in progress}}.

\end{thebibliography}

\end{document}